# AN EXPERIENCE-BASED EVALUATION PROCESS FOR ERP BIDS

Adnan Al Bar[1], Victor Basili[2], Wajdi Al Jedaibi[3] and Abdul Jawad Chaudhry[4]

[1]Department of Information Systems, Faculty of Computing & Information Technology,
King Abdulaziz University, Jeddah, Saudi Arabia
ambar@kau.edu.sa

[2]University of Maryland & Fraunhofer Center Maryland, College Park, Maryland, USA
King Abdulaziz University, Jeddah, Saudi Arabia
basili@cs.umd.edu

[3]Department of Computer Science, Faculty of Computing & Information Technology,
King Abdulaziz University, Jeddah, Saudi Arabia
waljedaibi@kau.edu.sa

[4]E-Management Department, King Abdulaziz University, Jeddah, Saudi Arabia
achaudhry@kau.edu.sa

## ABSTRACT

*Enterprise Resource Planning (ERP) systems integrate information across an entire organization that automate core activities such as finance/accounting, human resources, manufacturing, production and supply chain management… etc. to facilitate an integrated centralized system and rapid decision making– resulting in cost reduction, greater planning, and increased control. Many organizations are updating their current management information systems with ERP systems. This is not a trivial task. They have to identify the organization's objectives and satisfy a myriad of stakeholders. They have to understand what business processes they have, how they can be improved, and what particular systems would best suit their needs. They have to understand how an ERP system is built; it involves the modification of an existing system with its own set of business rules. Deciding what to ask for and how to select the best option is a very complex operation and there is limited experience with this type of contracting in organizations. In this paper we discuss a particular experience with contracting out an ERP system, provide some lessons learned, and offer suggestions in how the RFP and bid selection processes could have been improved.*

## KEYWORDS

*Enterprise Resource Planning (ERP), Request for Proposal for ERP (ERP RFP), ERP Contracting, Business Process, ERP Implementation, ERP Proposal*

## 1. INTRODUCTION

In this paper we present our experience with contracting out an ERP system at King Abdulaziz University in Saudi Arabia. We describe the process that was applied, the consequences of various decisions, and what we believe could have been done to improve the process.

The goal is to present a set of lessons learned, specifically identifying the high risk areas involved in writing a RFP, selecting a contractor, and setting up for the implementation. Based upon our experience, we will suggest a process that we believe would have shortened the delivery time, reduced costs, and provided a better final product. The goal of this paper is to help organizations, particularly in the public sector, be aware of the pitfalls involved in an ERP implementation and provide recommendations on how to avoid these risks.

King Abdulaziz University (KAU) was established in 1967 as a national university aimed at spreading higher education in the western area of Saudi Arabia. It offers university education to both female and male students [1]. The university has witnessed much improvement in quality and quantity since it was first established, becoming one of the more distinguished universities in terms of the number of students, the number of scientific and theoretical fields of study, and the quality of its programs. It is also the only university in Saudi Arabia that offers certain specializations such as Sea Sciences, Geology, Nuclear Engineering, Medical Engineering, Meteorology, Aviation, and Mineralogy.

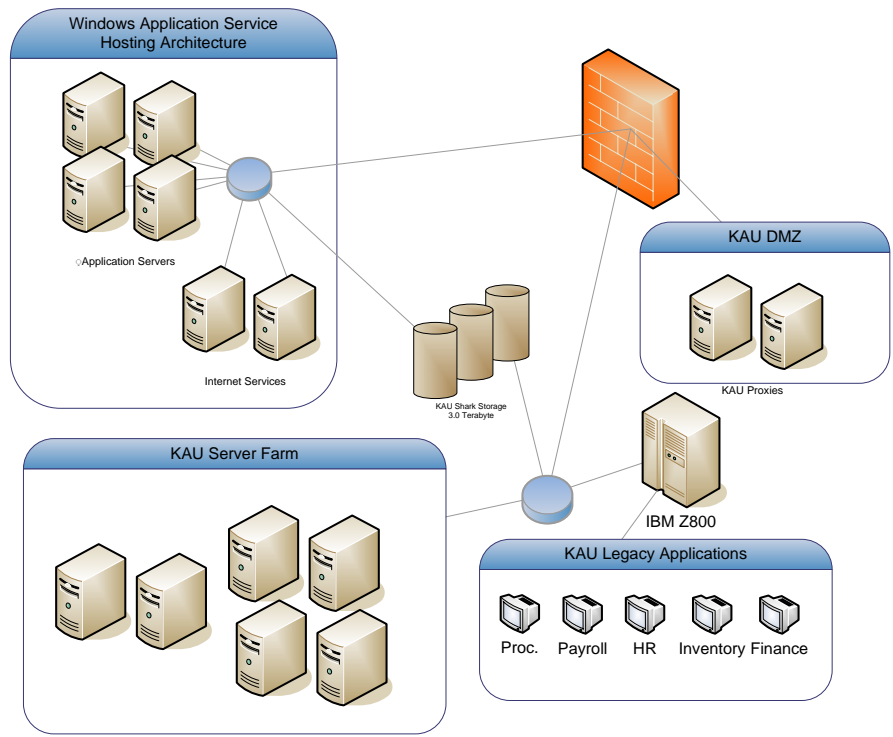

**Figure 1. KAU computing architecture showing legacy applications**

Early on, KAU adopted and developed many information systems to automate the administrative work that allowed different departments to execute their transactions through the systems, e.g., Human Resource, Finance, Budget and Planning (Accounting), Purchasing, Contracts, Warehouse, Projects, etc. But it became clear that recording day-to-day activities and monitoring the work within the department wasn't sufficient. These legacy systems were developed based on third generation programming language COBOL/CICS and using a DB2 database residing on an IBM mainframe (IBM-Z series series ). Components were developed separately over several years with minimal integration among the components and many recognized problems, such as incomplete business functionalities, duplicated business process, data redundancy due to inconsistent and un-normalized databases, no real automation due to the lack of integration, data inconsistency and inaccuracy. The system was only used as data storage as little or no business logic was being applied.

Figure 1 depicts KAU system landscape including KAU legacy application systems. The computing environment there, under a continuous demand for more services, has grown vastly to include many servers for Windows hosting architecture and internal/external domain services. Microsoft Active Directory is an example of a much needed service by many applications today at KAU computing services, since it facilitates a single sign-on service for KAU'ers. In addition, more and more web-based application appeared to the scene as depicted creating a group of diverse applications and services. This diversification of services resulted in creating a huge server farm and KAU Demilitarized Zone (DMZ) along with the IBM COBOL/CICS legacy systems. The original database is IBM DB/2, however, with the advent of other applications and platforms many other databases are in use today. It was clear there was a clear need to develop a fully automated, integrated system that facilitated the flow of all business activities through all departments, from start to finish in one single system with all the needed data shared.

Enterprise resource planning (ERP) systems integrate the internal and external management of information across an entire organization, embracing Finance/Accounting, Human Resources and Procurement, Contract Management, etc. ERP systems automate this activity with an integrated software application. The purpose of ERP is to facilitate the flow of information among all business functions inside the boundaries of the organization and to manage the connections to outside stakeholders [16].

With the development of a Saudi Arabian e-Government initiative [8, 9, 15], KAU bought into the vision of being a "Paper-Less University" and was the among the first public sector organizations in Saudi Arabia to implement an ERP system that would enable a fully automated system internally (by integrating all departments in one system) and externally (by adopting capabilities that ensure integration with other public sectors). The University decided to implement an ERP system and created an E-management department reporting directly to the vice president for administrative affairs. Amongst the authors of this paper is the head of the E-Management department, who was the supervisor of KAU SAP implementation project. Figure 1 shows SAP enterprise system architecture which has been adopted at KAU SAP project. In Figure 2 we show the real KAU SAP system architecture landscape. This landscape reside on an IBM system Z mainframe. The landscape is further subdivided into production and quality/development environment.

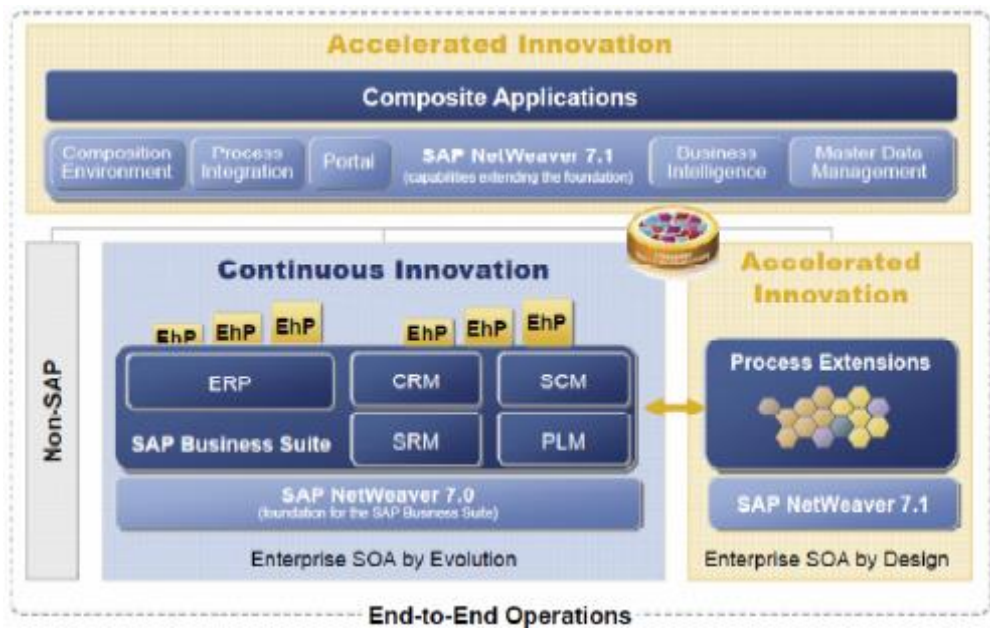

Figure 1 SAP Enterprise System Architecture (adopted from SAP)

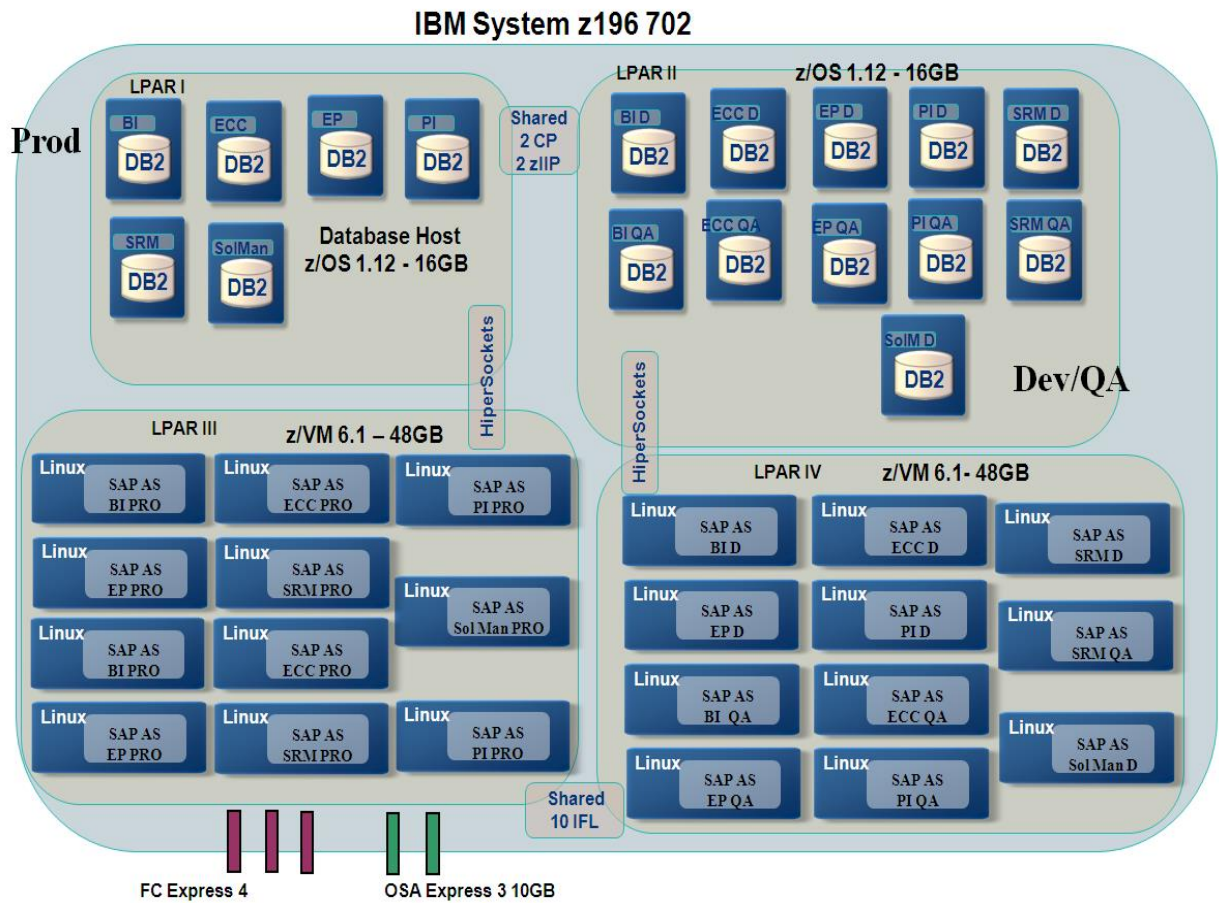

Figure 2 KAU SAP System Landscape Architecture

## 2. RELATED EXPERIENCES

In reviewing the literature [2, 3, 4, 5, 6, 7] there were several articles that discussed their lessons learned from the implementation of their ERP system. Although there were none that we could find that specifically addressed their experience with writing an RFP and selecting a bid, there were lessons identified that certainly were relevant to the topic. One of the studies, *Bondarouk* and *Van Riemsdijk* [2] suggested that a six-month investigation period was important to get ready for the implementation. *Dieringer* [3] recommended that selection an ERP package requires a lot of homework beyond a demonstration of its capabilities. They implied that a universal buy-in is needed, starting with the top executives down to everyone in the organization.

There was a consistent theme throughout all the papers that a redesign of the business processes were critical, i.e., the goal of an ERP is not to automate what the organization is already doing but to improve its processes, which involves reengineering the processes and moving closer to the processes offered by the particular ERP system being selected. *Lykkegaard* and *Gemela* [5] make the point that the organizations adapting an ERP system must be prepared for a transformational change and not treat it as a regular IT project. An important goal here is to minimize the number of customizations as it is the customizations that cause unanticipated costs, schedule delays, and a higher likelihood of bugs. This involves lots of discussion and acceptance and training for the stakeholders, the whole process of integrating the internal and external management of information across an entire organization is meant to be a learning experience. There is continual learning involved as the system evolves. However, we believe that if a business architecture of KAU business functions and their correspondent department and the business processes used was conducted and built, it would had support the transformational change that was clearly needed during various implementation phases of the project.

## 3. THE RFP DEVELOPMENT PROCESS

Figure 3 depicts the stakeholders of the RFP creation process and related intersections among them. In our case the executive management team was the University President and all of his deputies. Business process owners are the managers of the administrative departments such as the University Human Resource Managers and the Financial Manager. The IT deanship at the University represented the development and infrastructure stakeholders. The most complicated area for settlement is noted in red, where all classes of stakeholders must agree on the specific details. Decisions made in these ternary intersections must be based on consensus; simply voting on issues would not be practical due to the fact that all stakeholders will be greatly affected by decisions made. The role of executive management is to ensure alignment of project goals and enforce a unified decision process. Business process owners are comprised of subject matter experts (SME), data owners, and process owners. They are the responsible personnel who have the authority to lead and direct the execution of transactions within their respective departments. Development and infrastructure groups represent the e-management, implementation team, and IT team.

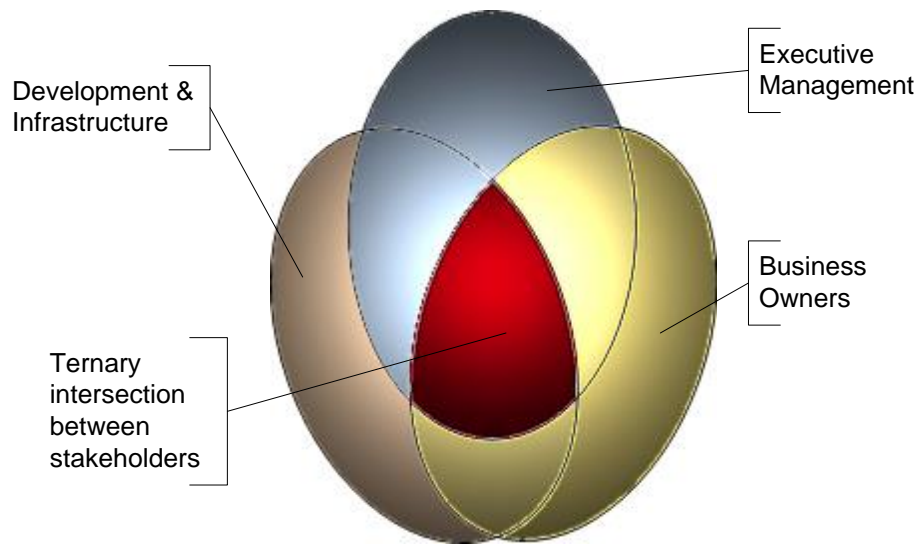

Figure 3. RFP Stakeholders

Contracting for an ERP implementation is difficult because there is such a large number of stakeholders, all of which need to learn over time, be heard, trained, negotiated with, and be in constant communication with each other. Business process owners need to be able to agree with and transition to reengineered processes. This involves a learning curve and a form of continual evolution to the Should-Be business processes. It is natural for some business owners to fear that the evolved system will not support them in their activities. It also means change from what they are currently doing.

The implementation of an ERP is not a typical development process with a standard set of requirements, design, and code but a process of customizing the code from an existing system. There is a general lack of knowledge and experience with this type of implementation process as well as with understanding the bounds and limits of the available ERP systems.

Added complexities come from the fact that KAU is a public sector organization that must communicate with the larger e-government organizations. This means that the standard government-based policies and processes are different from the ERP standard activities for-profit businesses and so the mapping to the customized set of practices can be more complex. It also means that there is another external major stakeholder and system, the e-government system YESSER, with which the KAU stakeholders and system need to interact with. .Once the decision had been made to implement an ERP system, the goal was to do it as quickly as possible. In retrospect, it is clear that the development of the RFP and Bid selection were too rushed. The whole process took about 9 months from start to announcing the winner. Of this, ten weeks was for the ERP readiness report and 6 weeks for the RFP preparation. Due to government business rules that require the RFP to be in the newspaper for 60 days in order for the bidder to respond to the RFP and reply with their proposal. It took us another 12 weeks to evaluate the proposals and select the winning bid. In the following section we will discuss each step of the RFP development process in more details.

### 3.1. ERP Readiness Assessment

As a start, the e-management team conducted several meetings with all stakeholders to decide on the appropriate approach for the road ahead. After reviewing the literature and what were considered the industry best practices in ERP system implementation [13], the e-Management team began to analyze the university's requirements by conducting several internal workshops to identify a list of business processes from all the relevant departments. It was decided to conduct an ERP readiness assessment [7] to help characterize KAU's maturity with respect to three important factors, (1) Organization (people), (2) Business Process, and (3) IT infrastructure and decide on the road to take. Three external experts were hired to conduct the readiness assessment and provide support in writing the RFP. Two had experiences in implementing ERP systems and the third was hired to assist with the IT infrastructure. The readiness analysis of KAU's business processes included a comparison of the required functionality against the features offered by the major ERP systems and an analysis the availability of skilled resources in the region. It came up with four options. The first option (continue with current system) was ruled out due to the fact that the current systems lack the necessary functionality to support the current business intellectual properties and future expansion of the business needs. The second option (Custom Built) was also not a viable option as the university core competency is to offer education and not to build systems. The third option was to select an ERP package. The fourth option (Hybrid) was to combine an ERP with internally developed systems was also rejected as it would not be able to scale to the KAU's future business growth. It was concluded that selecting an ERP package (option 3) from a world-class vendor would be the most viable option for KAU. The analysis strongly suggested that the ERP roll out be staggered over a period of time, starting with the Finance and Accounting module and introducing other departments according to a roll-out readiness map. It was felt that a multi-phased implementation as opposed to a "Big-Bang" approach would minimize the risk of failure, allowing the organization to learn from its mistakes and gradually build up the in-house talent pool and sponsorship that would provide a grass-roots "Buy-In" for both the IT organization and the Business divisions.

Based upon the readiness report and the requirements gathered, the e-Management team developed an RFP (Request for Proposal) which included the Scope of the Project and published a Public Bid to select an implementer who would be capable of implementing an ERP system for a public sector organization. In order to satisfy the business process owners and mitigate their fears of moving to a new system, the RFP business process requirements were specified as 'as-is' requirements, duplicating what was currently being done.

### 3.2. Business Processes Tradeoffs

As stated earlier, ERP systems are not developed based upon a set of requirements but mapped into a predefined collection of components that offer a limited set of variations on a set of business processes. This mapping method requires a huge effort where the current business processes have to be discussed (As-Is Business Processes), and evolved (Should-be Business Processes) to improve the way the business processes satisfy the organization's needs, and then compared with business processes offered by the ERP. This approach should necessarily lead to a form of Gap Analysis during bid evaluation. The real time and effort expended on the implementation comes from the customization of those components to meet the needs of the organization's should-be business processes.

The ultimate goal in selecting an appropriate ERP system is to identify each particular business process offered by a particular ERP system, understand the limitations of the ERP Standard Processes with respect to the Should-Be business processes, and propose a solution that minimizes the amount and cost of customization, i.e., that enables the organization to adapt their Should-Be business processes with mostly minor changes. When we require changes to the existing set of components, each upgrade of the current version of the ERP system, will require these changes to be made again to the next version of the system to be consistent with the next release of the ERP system.

Based upon the trade-offs between what might not make it to the should-be processes and the lack of sufficient knowledge about the various ERP offerings, the decision was made to create an RFP that was general in nature aimed at leaving open possible issues to be settled during implementation. This was due mostly to the lack of time to reach a consensus on the should-be processes. As it was not possible to predict all the possible implications of the integrated set of business process, it was decided to leave that to the particular ERP that was chosen. In retrospect, this was a major mistake. The too general nature of the RFP was evidenced by the fact that the bids varied in price by an order of magnitude.

### 3.3. ERP Limitations

Due to the lack of team experience with respect to an ERP implementation, and the number of limitations in the collection of ERP systems, many of these limitations were not well understood by the KAU team. For example, the ERP systems are not developed based upon a set of requirements but mapped into a preset collection of components that offer a limited set of variations on a set of business processes. As stated above, the real time and effort comes from the customization of those components to the needs of the organization's business processes. In retrospect, it was clear that a detailed *gap analysis* needed to be performed identifying the kinds of changes that needed to be made to each of the proposed ERP systems and categorized by their level of complexity.

Each ERP has limits in terms of what can be handled. Most limitations of ERP packages can be categorized into functional and technical limitations. Functional limitations are mismatches between the ERP standard process and the organization's business process and the organizations business process cannot be implemented without a prohibitive amount of effort.. An example of a functional limitation is that some of the standard business processes of ERP systems cannot be automated through an approval workflow, e.g., a required approval by a manager. But the KAU business processes require a full approval business process workflow, more than five levels of approval. Also, some business rule validation has to be hardcoded into the existing set of ERP components because it cannot be made a variation of any of the existing business rules. This is a major drawback with respect to system upgrades as the hardcoded business rule have to be handled with care otherwise they will be overridden by the upgrade.

Technical limitations are encoded restrictions in the ERP system design that are very difficult to overcome. As an example of a technical limitation, SAP cannot handle name text fields larger than 40 characters. This turned out to be a problem for KAU's standard terminology for business entities. This limit is actually hardcoded in SAP, i.e., it is not a parameter. Another example of a technical limitation is that the lack of support to configure the financial fiscal year according to your local calendar as in our case for the Hijri calendar, which is the official calendar of Saudi Arabia.

As implied earlier, an important factor causing problems was that the KAU ERP was for a Saudi government agency and must adhere to government rule restrictions, real or imagined, on their business processes, i.e. going from As-Is to Should-Be involved complicated negotiations. For example, during the negotiation about improving the current set of business processes with the business process owner, there was resistance encountered. The Business process owner would object to the change on the basis that he thought the current rules were required by the Ministry of Finance and can't be changed. It was also a painful exercise to resolve conflict among business owners with regards to some file names in shared screens with each owner insisting on keeping their own familiar terminology. In ERP packages, field names are usually defined once and used everywhere. KAU business process owners came from a COBOL legacy system world, where every application has its own terminology and classification encodings. It was evident that letting go of old processes and encodings was difficult for them. In this regard, the whole ERP project was viewed as a threat rather than improvement.

Another serious issue was that the RFP did not specify the common browsers used by the business process owners and the wining RFP did not support the use of that browser. The most common browser at KAU was Google chrome and the winning bid only supported Internet Explorer version 8 at the time of the project. This created serious concerns from the business process owners. Unfortunately, the business process owners did not have an opportunity to review the RFP, even though it is not clear they would have thought about specifying the browser.

## 4. USING LESSONS LEARNED TO PROPOSE A BETTER RFP DEVELOPMENT PROCESS

A lot of unanticipated cost, effort and time were expended because of the lack of an available functional and technical ERP consultant at KAU (and possibly anywhere) for developing an RFP to contract for an ERP system. Based upon what happened we would propose a different approach to adopt an ERP for an enterprise organization like KAU and a different approach for bid selections, too. In our discussions with successful implementers of ERP systems, it appears to be quite common that the successful implementation was not the initial attempt at an implementation. The highest level change would be to spend more time and money upfront, e.g., have a separate activity (e.g., a contract) for developing the RFP and evaluating the bids so there would be a minimum of surprises during the implementation of the ERP. As others in the literature have suggested [10, 11], we should have spent more time up front identifying what needed to be done and to learn more about ERP implementation and more of its hidden pitfalls. One approach for starting the process would have been to hire (or contract) a team that has the right experience so that KAU personnel work with them directly to gain the experience needed [12].

### 4.1. Identifying and specifying goals for the ERP

We now know that from the early start of the project, the goals were not clear to all stakeholders. In fact, in some areas conflicting goals existed! While the e-management team knew to some extent what was to come, other stakeholders did not. For example, business process owners thought that it was just another application and the change of the actual processes would be minimal. That was, however, not the case. Reactions differed when integration of the systems was to take place. For instance, integration means a unified encoding system must be used; hence specific locally exercised flexibility must be abandoned. Any false or wrong piece of data entered will be flagged if not by the system itself then by the users of the other integrated modules of the ERP. As the project progressed, the goals of each group of the stakeholders became clearer and more conflicting. Thus, more resistance came to the surface.

If we were to do this again, we would include a couple of opening workshops, e.g., Joint Application Design sessions [17], where we would discuss the goals, e.g., improving the business processes so they could be more efficient, automated, and integrated with each other to provide a seamless system, bounding the changes that could be made, etc. We could then discuss the strategies for achieving these goals. All stakeholders would be required to participate in these workshop and all concerns would be aired and a consensus drawn. A key factor is that there be an executive sponsor, the project owner, who is high enough in the organization to be able to make the necessary final decisions.

### 4.2. Business Process Development

Based upon the agreed goals, we would perform a thorough analysis of how well KAU business processes are performing. Such an analysis would bring hard evidence to the discussion with business process owners on the need to change various processes and which current processes are not optimal. This discussion is needed to make it clear to the business process owners what processes should be changed in order to achieve the stated goals in order to yield an overall improvement of the internal and external management of information across the entire organization. Given the As-Is, what should be the To-Be. This requires a balancing of what would be an improvement to the processes that would satisfy the goals and what would be acceptable to the current business process owners. One approach that would help the business process owners envision the potential improvements is to build prototype screens that show the new business processes and allow the business process owners to play with them

We would evolve these changes via a series of workshops where we would discuss how to improve the business processes, identify the hurdles in terms of discussions of the effect of changes in the business processes, the implications of integrating the systems, etc. All stakeholders would be required to participate in the workshops and all concerns would be aired and consensuses drawn. Ultimately, the executive sponsor would arbitrate the final set of decisions. It would be good for the workshops to occur over several days so that stakeholders would have the chance to think over the implications of each day's activities.

### 4.3. Improving Communication

A fundamental principle that should be emphasized during the entire process is to improve communication among all stakeholders. At any stage of the project, they should be made aware of the status of the project and all project related decisions. This information should even be disseminated to them periodically. They should be allowed to comment and express concerns enabling them to engage actively in the project. Having an electronic communication channel, e.g., group forums, would encourage this kind of interaction. In addition, we suggest that regular workshops and team meetings should be conducted, especially before deciding on major issues. This will facilitate communication and demonstrate to stakeholders that they are an integral part of the team.

### 4.4. Incorporating Continuous Learning

The results of these three activities would help create an organizational mindset that developing an ERP is a continuous learning activity. That is, as an organization we are evolving the ERP, learning from each decision we have made along the way. Because of the lack of internal experience, acceptance of the idea that we are learning step by step is important. It was a mistake to develop an RFP for the entire set of business processes without integrating a learning process into it. For example, it would have been useful to start the implementation with a particular set of business processes, e.g., human resources, and learn from them for implementing the next of processes. What is learned from that particular task could be used to do a better job of implementing the next set of business processes. It would allow the KAU team to build experience in each of the tasks and be better able to play a significant role in the implementation.

### 4.5. Developing the RFP

At this point we should be ready to develop the RFP. As stated above, it might be appropriate to build the system incrementally and select a first set of business processes for implementation. Since the contracting for an ERP is not contracting for a development but a modification of an existing system, it is clear that we would need to understand the differences between the specified business processes and those available from the various ERP systems in order to select the system that requires minimum modifications. It is the cost of modification that will be the major cost of the system. To this end, we should perform a gap analysis between what the various ERPs offer and what we need to satisfy our goals.

### 4.6. Gap Analysis

The bidding process required each bidder to demonstrate how their system would address various types of high level business processes, e.g., human resources. But we did not provide specific and detailed processes, such as the processes for hiring new employees of various types. As a government agency, KAU has a large number of complex hiring processes for different employee roles. It was not clear what kinds of configurations and customizations were needed. Configuration is the process of setting up the ERP system by selecting standard functionalities from a list of those supported by the system. Customization is the process of mapping ERP to organization's business process which might require code changes to create functionality that is not available through configuration.

A better approach would have been to provide a representative set of specific processes for each of the high level processes that would allow a more detailed Gap analysis.

### 4.7 The bid Evaluation process

As stated earlier the bids varied in cost by an order of magnitude. The rules were to accept the bid with the best fit solution at the lowest price. Unfortunately there was not enough information to make that judgment. This was due to lack of details in the RFP and to the lack of experienced ERP consultants at KAU when evaluating bids A session was held with business owners to ask for their feedback and get a review with them about the general requirements. However, the main goal was to spread the awareness among business owners that we were going for an ERP solution. It was not meant to solicit detailed comments about the RFP from the business process owners. The bid selection was viewed as a technical task. In retrospect this was a mistake not to solicit feedback as the process could have been used to create a more collaborative atmosphere and an educational experience for the business process owners.

As an attempt to better understand what bidders had to offer, they were asked to provide a live demonstration of the ERP system they were proposing. Some of the demos were scheduled for three full days. Bidders were given a scenario of a business process but it was not clear from the demo what was needed to customize the ERP as we didn't have a functional consultant to pin point issues related to the products offered and whether they suited KAU business processes. This limited our ability to make informed decisions about each vendor.

In retrospect, we should have done a mapping of each KAU business process for each bid to understand and assess what standard capabilities the bid offered. This was done at a high level, but not in sufficient details to make clear the customization and effort needed for that customization. For example, we did it for Human Resources in general but not at a detailed level, such as the hiring process.

One approach would be to build a table for evaluating the level of customization covering sufficient detail in the processes to make clear what customization would be needed (see Table 1).

For example, if each row represents a particular detailed sample business process and each column represents an evaluation of how that proposed ERP system would be rated in terms of a specific business process being available as standard (0), customization (1). This would give us significantly more information as to which proposed system would require the minimum amount of effort to implement and reduce the amount of customizations. This step should be done in advance of contracting the project.

Table 1. Bid Evaluation

| Business Processes | Bidders | | |
|---|---|---|---|
| | Bidder 1 | Bidder 2 | Bidder 3 |
| New Hiring | 0 | 0 | 0 |
| Permanent | 0 | 1 | 1 |
| Contract | 1 | 0 | 0 |
| Academic Hiring | 0 | 1 | 1 |
| Administrative Hiring | 1 | 0 | 0 |
| Hiring by Transfer | 1 | 0 | 1 |
| Rehiring | 1 | 1 | 1 |
| Retirement | 0 | 0 | 0 |
| Recruitment | 0 | 0 | 1 |
| Conference Attendance | 1 | 1 | 1 |
| Training Attendance | 1 | 0 | 0 |
| Performance Evaluation | 0 | 0 | 1 |
| Academic Promotion | 1 | 1 | 0 |
| Administrative Promotion | 0 | 0 | 1 |
| **Total** | **7** | **5** | **8** |

Thus, for this set of business processes, Bidder 2 would require the least effort. The results would also provide feedback on those business processes which would be the most trouble to implement and would give KAU the opportunity to see if those processes could be modified to minimize enhancements and developments. Based upon our principle of good communication, we would again run workshops with the relevant business process owners to see what adaptations could be accepted.

Moreover, similar to the gap analysis of business processes, it would be important to list all reports, forms and printouts in another table to evaluate the capability of each proposal and how their ERP system will address these requirements. It could be the case that the required format by business owners doesn't map to the standard reports/forms or does not exist; these kinds of formats can be customized which should be part of the evaluation process of the bid.

### 4.8. Identifying Technical Issues

Aside from identifying the gaps in the implementation of the business processes, an examination of the data migration and cleansing processes needs to be performed. Data migration and cleansing are major activities in an ERP implementation project. It is important for a data architect to be able to map the scattered existing data in different formats from the various existing legacy systems to a common, centralized, consistent master data file in the ERP implementation. Thus it is important to have a highly capable data architect as part of the KAU team. This cleansing involves identifying and effectively using master data from the KAU data resources such as legacy databases and archived HR files. The RFP needs to specify as part of the project, the identification and collection of the missing data from the various sources. A fair sample of current data should been made available to bidders flagging the amount of work needs to be accomplished in the transition.

### 4.9. Selecting the Bidder

At this point, armed with the results of the gap analysis and the identification of the various technical issues that need to be overcome, the bidder with the minimum number of customizations and the best technical support can be selected. The organization, at this stage, should be able to develop reasonable estimates of the cost and schedule and for evaluating the bids with respect to these parameters.

## 5. CONCLUSIONS

Contracting for an ERP system is a complex and difficult task, especially without having any previous ERP implementation experience. It involves (1) writing an RFP for a system that will be developed based upon the configuration and customization of an existing system and (2) selecting a bidder and proposed system based upon how closely their proposal minimizes the amount of customization. The writing of the requirements for such a system needs to be a compromise between what business processes already exist, what business process are available in the ERP system, and what business process would improve the way an organization does its business. There has been little discussed in the literature about the contracting process, which sets the stage for the implementation. In the KAU case, it is clear that not enough time was spent up front in marshaling all the stakeholders, focusing the RFP on the major issues, and evaluating the bidders appropriately. In many ways, the authors believe these activities are the key factors for implementing an ERP. Had they be done right, the actual implementation process would have been cheaper and faster.

Based upon our experience, this paper suggests an approach which should improve an organization's control of the project and allow it to minimize the risk of surprises during the implementation which should lead to a successful implementation.

### ACKNOWLEDGEMENTS

This project was funded by the Deanship of Scientific Research (DSR), King Abdulaziz University, Jeddah, Saudi Arabia. The authors acknowledge and extend their gratitude towards DSR for its technical and financial support.

**Authors**

[1]Adnan A. Al Bar

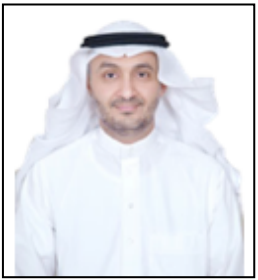

Adnan Albar is Assistant Professor of Information systems at the King Abdulaziz University. He holds a PhD degree in Computer Science from George Washington University. He served as chairman of the Information Systems department for two year and then as the Vice Dean for Development of Student Affairs. He supervised the implementation of SAP ERP project at the King AdulAziz University. He is a member of the ACM and IEEE. His main research interests are enterprise information systems, IT governance, and enterprise architecture.

[2]Victor R. Basili

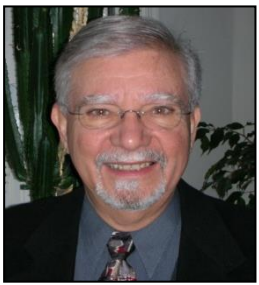

Victor Basili is Professor Emeritus of Computer Science at the University of Maryland. He holds a PH.D. in Computer Science from the University of Texas, Austin and two honorary degrees. He served as founding director of the Fraunhofer Center - Maryland and the Software Engineering Laboratory at NASA/GSFC. He serves as co-Editor-in-Chief of the Journal of Empirical Software Engineering, and is an IEEE and ACM Fellow. He works on measuring, evaluating, and improving the software development process and product.  For more information, please see http://www.cs.umd.edu/~basili/

[3]Wajdi Al Jedaibi

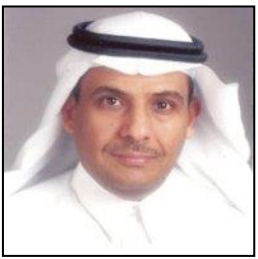

Wajdi Al Jedaibi is a faculty member of the Computer Science department in the Faculty of Computing & Information Technology at King Abdulaziz University. He served as the KAU IT Manager and then Dean of Information Technology at KAU. Wajdi was awarded the Ph.D in Information Technology – Software Engineering and MSc. in Software Systems Engineering both from George Mason University. His current research interests are: component-based software engineering, software measurement, and Open Source ERP systems.

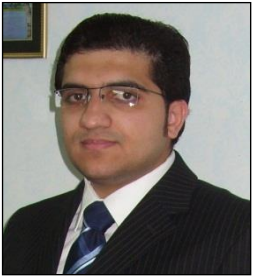

[4]Abdul Jawad M. Chaudhry

Professional system analyst and ERP Consultant, he has participated in several ERP implementation projects for both private and public (government) organizations, He received a professional associate certificate for SAP Consultation and was awarded appreciation certificates from leading organizations like: Saudi Arabian Airlines and IBM.